\pdfoutput=1

\documentclass[
    mathptm
    ,final            
  ]
  {aipproc}

\layoutstyle{6x9}

\newcommand{\nn}{\nonumber}
\def\stilde{\widetilde}

\pdfpagewidth=\paperwidth
\pdfpageheight=\paperheight

\usepackage{amssymb}

\begin{document}

\title{CP Violating Asymmetry in Stop Decay into Bottom and Chargino}



\author{Sebastian M. R. Frank}{
  address={Institute of High Energy Physics, Austrian Academy of Sciences, A-1050 Vienna, Austria},
}

\author{Helmut Eberl}{
  address={Institute of High Energy Physics, Austrian Academy of Sciences, A-1050 Vienna, Austria}
}

\begin{abstract}
In the MSSM with complex parameters, loop corrections to the decay of a stop into a bottom quark and a chargino can lead to a CP violating decay rate asymmetry.
We calculate this asymmetry at full one-loop level and perform a detailed numerical study, analyzing the dependence on the parameters and complex phases involved. In addition, we take the Yukawa couplings of the top and bottom quark running. We account for the constraints on the parameters coming from several experimental limits.
Asymmetries of several percent are obtained. We also comment on the feasibility of measuring this asymmetry at the LHC.
\end{abstract}

\maketitle


\section{Introduction}

Supersymmetric extensions of the SM can contain new sources of CP violation, which can lead to an explanation of the baryon asymmetry of the universe (BAU) via Electroweak Baryogenesis. If one chooses some of the MSSM parameters to be complex, processes can lead to CP violating asymmetries. But even if BAU cannot be explained, studies of CP violation and values of potentially complex parameters are important.

\section{Decay Rate Asymmetry $\delta^{CP}$}

We define the CP violating decay rate asymmetry of the decay $\tilde t_i \to b \, \tilde \chi^+_k$ as
{\setlength{\abovedisplayskip}{4pt plus 0pt minus 0pt}
\setlength{\abovedisplayshortskip}{4pt plus 0pt minus 0pt}
\setlength{\belowdisplayskip}{-7pt plus 0pt minus 0pt}
\setlength{\belowdisplayshortskip}{-7pt plus 0pt minus 0pt}
\begin{equation}
\delta^{CP} = ( \Gamma^+(\tilde t_i \to b \, \tilde \chi^+_k) - \Gamma^-(\tilde t^*_i \to \bar b \, \tilde \chi^{+ c}_k) ) /
( \Gamma^+(\tilde t_i \to b \, \tilde \chi^+_k) + \Gamma^-(\tilde t^*_i \to \bar b \, \tilde \chi^{+ c}_k) ) \, .
\label{eq:deltaCP}
\end{equation}
}\\
We write the one-loop decay widths as $\Gamma^\pm
\propto \sum_s |\mathcal{M}^\pm_\mathrm{tree}|^2 + 2 \mathrm{Re}
( \sum_s ( \mathcal{M}^\pm_\mathrm{tree} )^\dagger
\mathcal{M}^\pm_\mathrm{loop} )$.
Since there is no CP violation at tree-level and assuming that the one-loop contribution is small compared to the tree-level, the decay rate asymmetry can be approximated by
{\setlength{\abovedisplayskip}{3pt plus 0pt minus 0pt}
\setlength{\abovedisplayshortskip}{3pt plus 0pt minus 0pt}
\setlength{\belowdisplayskip}{-4pt plus 0pt minus 0pt}
\setlength{\belowdisplayshortskip}{-4pt plus 0pt minus 0pt}
\begin{equation}
\delta^{CP} \cong \frac{\Gamma^+ - \Gamma^-}{2 \Gamma_\mathrm{tree}} \nn = A^{CP}_+ - A^{CP}_- \quad , \quad
A^{CP}_\pm = \mathrm{Re} \Big( \sum_s ( \mathcal{M}^\pm_\mathrm{tree} )^\dagger
\mathcal{M}^\pm_\mathrm{loop} \Big) / \sum_s |\mathcal{M}_\mathrm{tree}|^2 \, .
\label{eq:deltaCP_approx_ACP}
\end{equation}
}\\
The matrix elements $\mathcal{M}^\pm_\mathrm{tree}$ and $\mathcal{M}^\pm_\mathrm{loop}$ are functions of the couplings $B^{R,L}_\pm$ and form factors $\delta B^{R,L}_\pm$, respectively. The explicit forms are given in~\cite{Frank2008}.
Introducing $C^{i j}_\pm = B^i_\mp \delta B^j_\pm$ ($i,j = R,L$) we can express $\mathcal{M}^\pm_\mathrm{loop}$ in terms of $\mathrm{Re} (C^{i j}_\pm)$ which have the structure
{\setlength{\abovedisplayskip}{1pt plus 0pt minus 0pt}
\setlength{\abovedisplayshortskip}{1pt plus 0pt minus 0pt}
\setlength{\belowdisplayskip}{-9pt plus 0pt minus 0pt}
\setlength{\belowdisplayshortskip}{-9pt plus 0pt minus 0pt}
\begin{displaymath}
\mathrm{Re} (C^{i j}_\pm) \propto \mathrm{Re} ((b g_0 g_1 g_2)^\pm \times \mathrm{PaVe}) = 
\mathrm{Re} (b g_0 g_1 g_2) \mathrm{Re} (\mathrm{PaVe}) \mp \mathrm{Im} (b g_0 g_1 g_2) \mathrm{Im} (\mathrm{PaVe})
\end{displaymath}
}\\
where $(b g_0 g_1 g_2)^-$ means complex conjugate, $b$ is the coupling at tree-level, $g_0 g_1 g_2$ are couplings of the three vertices
and PaVe stands for the Passarino-Veltman-Integrals. This leads to the decomposition into both CP
invariant and CP violating parts $\mathrm{Re} (C^{i j}_\pm) = C^{i j}_\mathrm{inv} \pm C^{i j}_\mathrm{CP}$ / 2
with the definitions $C^{i j}_\mathrm{inv} \propto \mathrm{Re} (b
g_0 g_1 g_2) \mathrm{Re} (\mathrm{PaVe})$ and $C^{i j}_\mathrm{CP}
\propto - 2 \mathrm{Im} (b g_0 g_1 g_2) \mathrm{Im}
(\mathrm{PaVe})$. We can see that we need not only the couplings
but also the PaVe's to be complex (i.e. at least a second decay channel kinematically open) in order to obtain a non-zero
$\delta^{CP}$. The asymmetry $\delta^{CP} = A^{CP}_+ - A^{CP}_-$ becomes
{\setlength{\abovedisplayskip}{2pt plus 0pt minus 0pt}
\setlength{\abovedisplayshortskip}{2pt plus 0pt minus 0pt}
\setlength{\belowdisplayskip}{-4pt plus 0pt minus 0pt}
\setlength{\belowdisplayshortskip}{-4pt plus 0pt minus 0pt}
\begin{displaymath}
\delta^{CP} = \Big( \Delta (C^{RR}_\mathrm{CP} + C^{LL}_\mathrm{CP}) -
2 m_b m_{\tilde \chi^+_k} (C^{RL}_\mathrm{CP} + C^{LR}_\mathrm{CP}) \Big) / \sum_s |\mathcal{M}_\mathrm{tree}|^2
\  , \ 
\Delta = (m_{\tilde t_i}^2 - m_b^2 - m_{\tilde \chi^+_k}^2)
\end{displaymath}
}\\
where one can neglect the second term due to the smallness of the bottom mass.

\section{CP Violating Contributions}

\begin{figure}[htbp]
\includegraphics[width=0.4\textwidth]{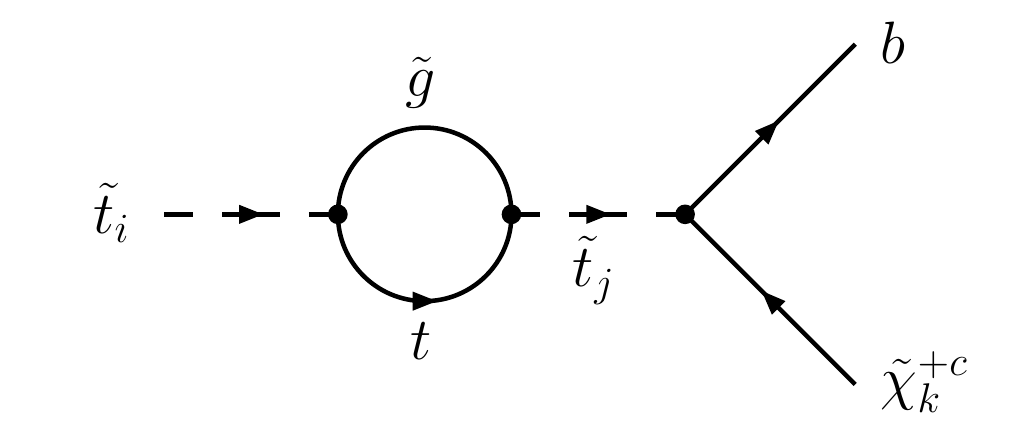}
\caption{Leading contribution of CP violation in $\tilde
t_i \to b \, \tilde \chi^+_k$ at one-loop level in the MSSM with
complex couplings ($i,j,k=1,2$).}
\label{fig:CP-violation}
\end{figure}

In principle, 47 one-loop diagrams can contribute. If the channel $\tilde t_i \to t \, \tilde g$ is kinematically open, the selfenergy graph in Fig.~\ref{fig:CP-violation} and the vertex graph with $\tilde g$ exchange should dominate due to the strong coupling. But numerically only the selfenergy graph dominates and we thus concentrate on this diagram.
\vspace{-7pt}

\noindent
The general matrix element can be written as ${\cal M}^+ = i \bar u(k_1) ( \delta B^R_+ P_R + \delta B^L_+ P_L ) v(-k_2)$.
The couplings as well as the Passarino-Veltman-Integrals are defined in~\cite{Frank2008}.
The form factor for the process is
{\setlength{\abovedisplayskip}{5pt plus 0pt minus 0pt}
\setlength{\abovedisplayshortskip}{5pt plus 0pt minus 0pt}
\setlength{\belowdisplayskip}{-6pt plus 0pt minus 0pt}
\setlength{\belowdisplayshortskip}{-6pt plus 0pt minus 0pt}
\begin{equation}
\delta B^R_+ = 2 m_{\tilde g} m_t B^R_{k j} ( G^{R *}_i G^{L}_j + G^{L *}_i G^{R}_j ) B_0 /
( (4 \pi)^2 (m_{\tilde t_i}^2 - m_{\tilde t_j}^2) )
\label{eq:delta_B_R-selfenergy}
\end{equation}
}\\
using the arguments $(m_{\tilde t_i}^2,m_{\tilde g}^2,m_t^2)$ for
the $B_0$-function. Note that $i \neq j$ in order to contribute to CP
violation.

\section{Numerical Results}

\begin{figure}[htbp]
\begin{tabular}{cc}
\includegraphics[width=0.4\textwidth]{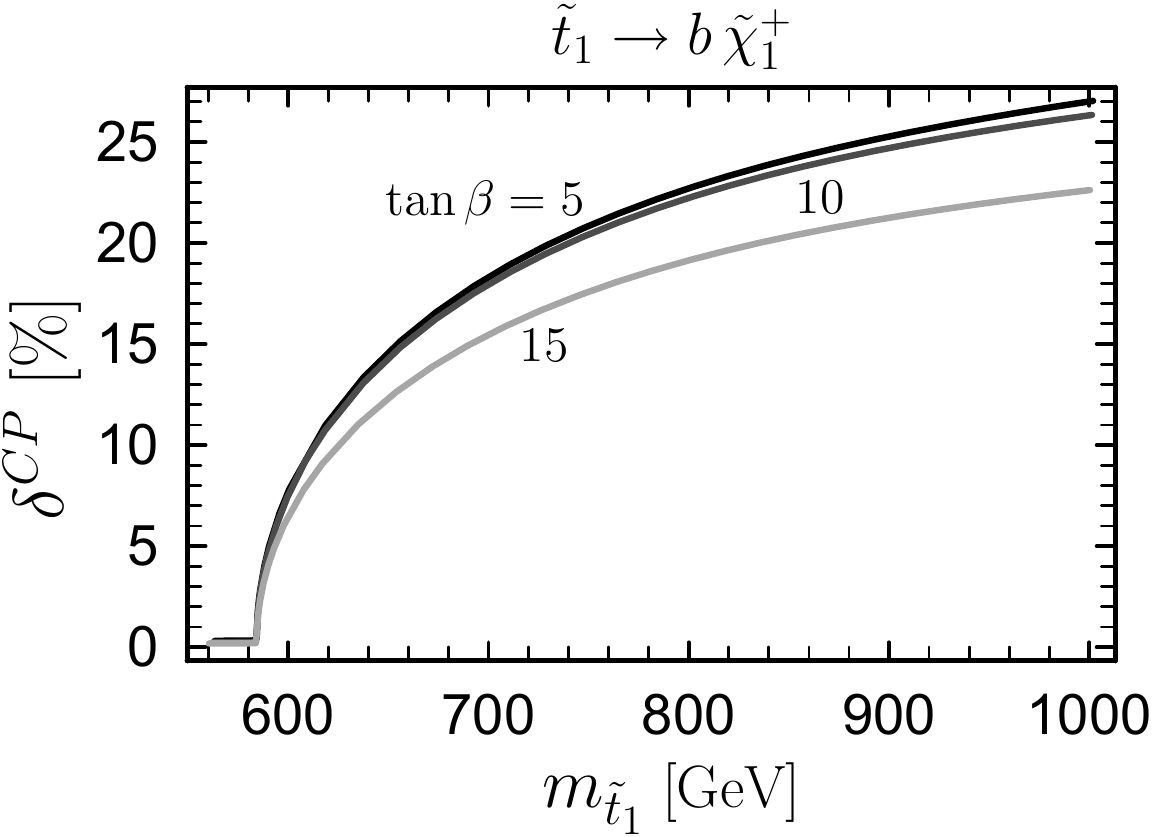} &
\includegraphics[width=0.4\textwidth]{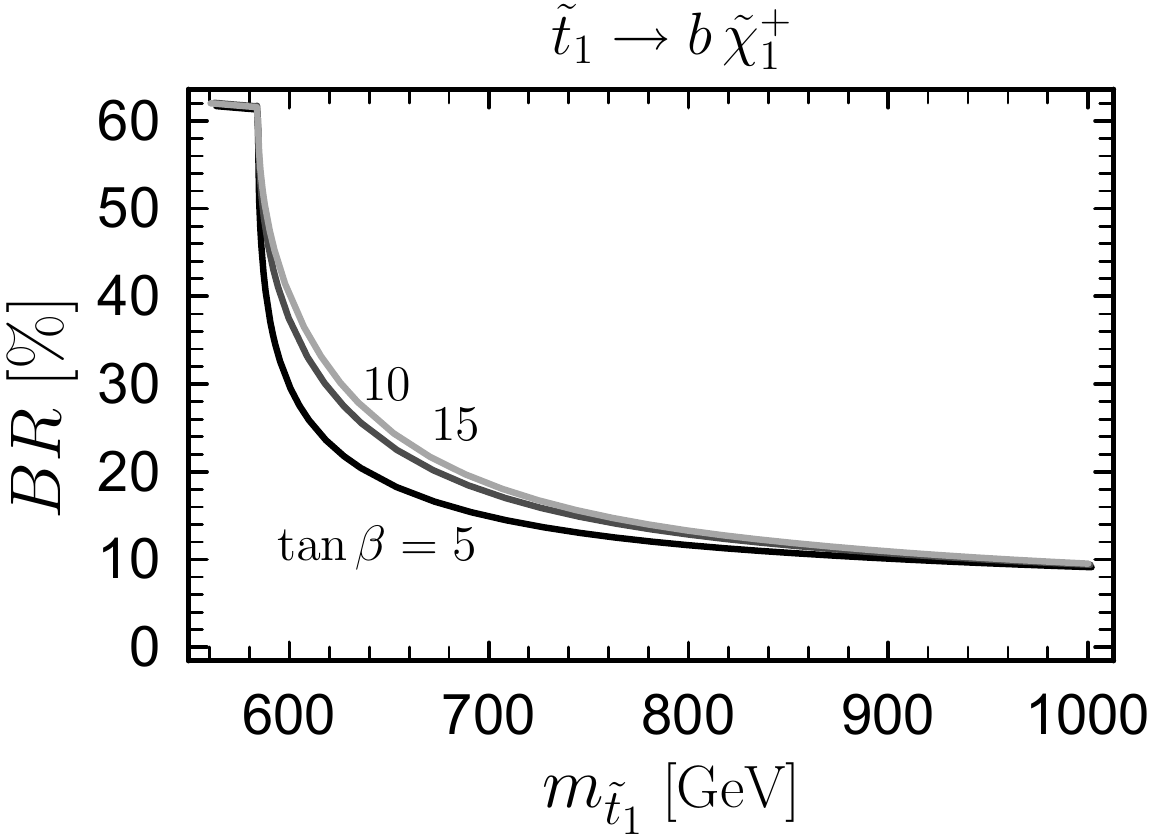}
\end{tabular}
\caption{$\delta^{CP}$ and $BR$ as a
function of $m_{\tilde t_1}$ ($M_{\tilde Q}$ varied) for various values of $\tan \beta$.}
\label{fig:delCP&BR_mstop1_TB_all}
\end{figure}

\begin{figure}[htbp]
\setlength{\unitlength}{1mm}
\begin{picture}(60,44)
\put(0,0){\includegraphics[width=0.4\textwidth]{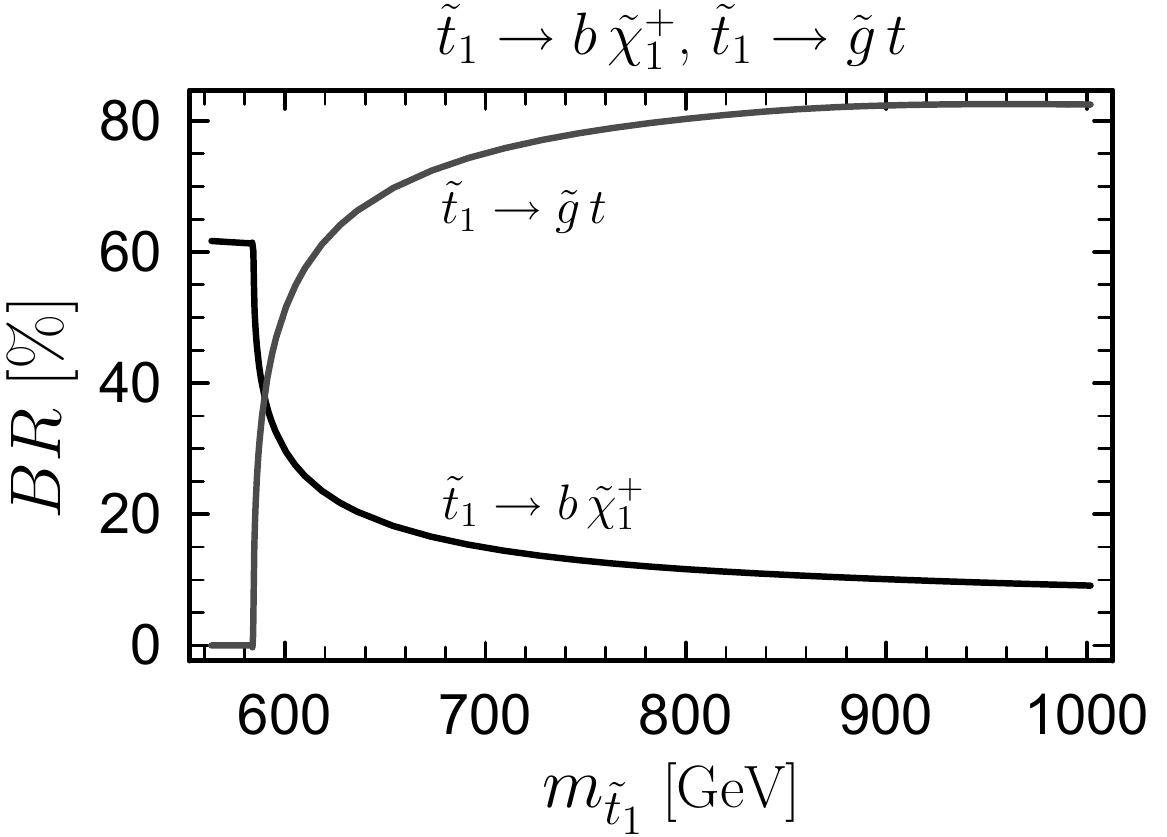}}
\put(0,40){\footnotesize (a)}
\end{picture}
\hspace{2mm}
\begin{picture}(60,44)
\put(0,0){\includegraphics[width=0.4\textwidth]{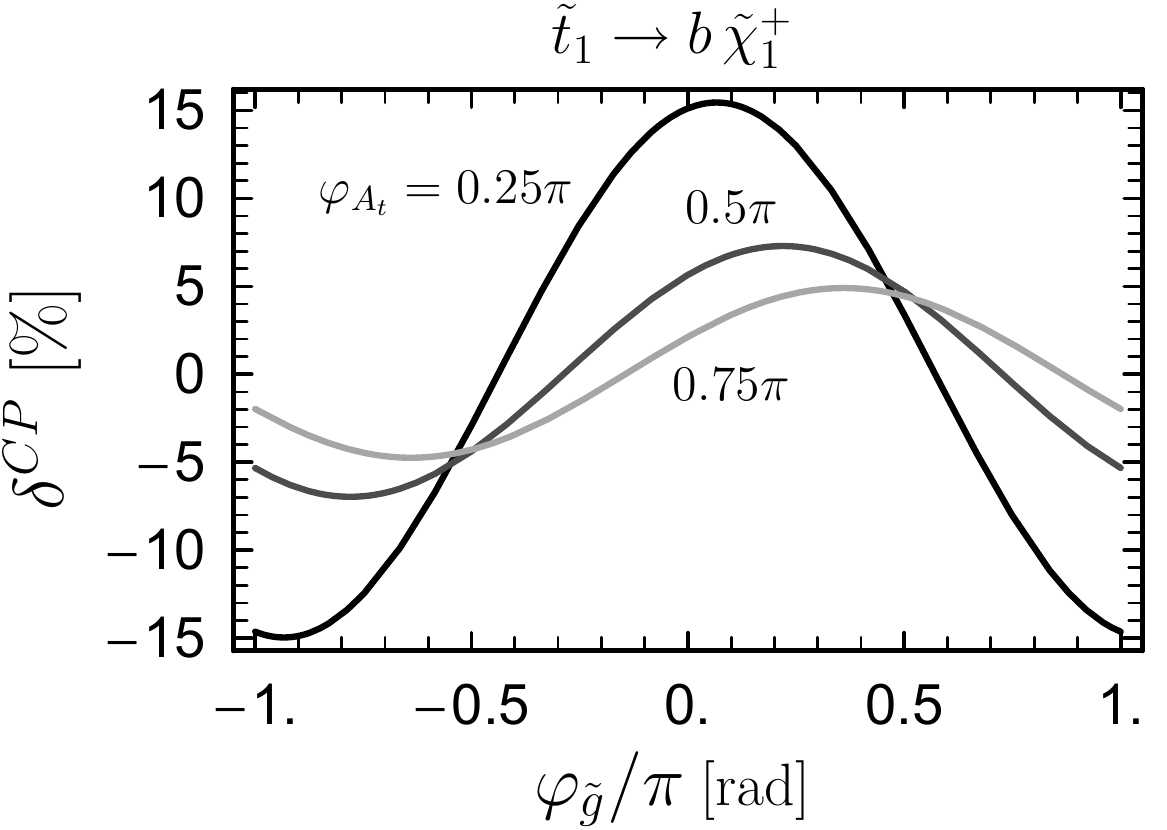}}
\put(0,40){\footnotesize (b)}
\end{picture}
\caption{(a) Comparison of $BR(\tilde t_1 \to b \, \stilde \chi^+_1)$ and $BR(\tilde t_1 \to \tilde g \, t)$ as a
function of $m_{\tilde t_1}$ ($M_{\tilde Q}$ varied).
\protect\phantom{(b)}
(b) $\delta^{CP}$ as a function of $\varphi_{\tilde g}$ for various $\varphi_{A_t}$.}
\label{fig:BR_mstop1_and_delCP_phiGluino_phiAt_only_gluino}
\end{figure}

We present numerical results for the decay rate asymmetry $\delta^{CP}$ as well as the branching ratio ($BR$) of the process $\tilde t_1 \to b \, \tilde \chi^+_1$.
The 47 contributions were calculated by using \mbox{\textsc{FeynArts}}~\cite{Hahn2001}. Furthermore, the two gluino graphs and others we calculated independently and also cross checked numerically. Our parameter points are consistent with constraints coming from the EDM of the electron (private code), $B \to X_s \, \gamma$ and the cold dark matter relic density (\mbox{\textsc{micrOMEGAs}}~\cite{Belanger2006}).\\
For the MSSM input parameters we take the 3rd generation SUSY breaking parameters $M_{\tilde Q}=M_{\bar u}=M_{\bar d}=650$~GeV and $M_{\tilde L}=M_{\bar e}=600$~GeV, $|A_t|=|A_b|=|A_{\tau}|=190$~GeV and the complex phases $\varphi_{A_t}=\varphi_{A_b}=\varphi_{A_{\tau}}=\pi /4$, $\varphi_{\mu}=0$, $\varphi_{M_1}=0$ and $\varphi_{\tilde g}=0$ (at first). We further set $M_2=150$~GeV, $|M_1|=M_2/2$, $|\mu|=830$~GeV, $\tan \beta=5$, and $M_{A^0}=1000$~GeV.
These parameters imply that $\varphi_{A_t}$ is at first the only source of CP violation, the chargino $\tilde \chi^+_1$ is gaugino-like, and $\tilde t_1$ and $\tilde t_2$ have a low mass splitting but high mixing.\\
In Fig.~\ref{fig:delCP&BR_mstop1_TB_all} we show $\delta^{CP}$ and $BR$ as a function of $m_{\tilde t_1}$ for $\tan \beta=5,10,15$. The parameter $m_{\tilde t_1}$ is shown for convenience, the parameter actually varied is $M_{\tilde Q}$ from $500$ to $1000$~GeV. One can see the threshold of the $\tilde t_1 \to \tilde g \, t$ decay at $\sim 583$~GeV, after which the gluino contributions account up to $\sim 98 \, \%$ to $\delta^{CP}$. However, if this decay channel opens, the $BR$ drops quickly.
This general feature and thus permanent conflict between $\delta^{CP}$ and $BR(\tilde t_1 \to b \, \stilde \chi^+_1)$ can also be seen in Fig.~\ref{fig:BR_mstop1_and_delCP_phiGluino_phiAt_only_gluino}a. If $\tilde t_1 \to \tilde g \, t$ is possible, $\delta^{CP}$ is large but $BR$ is small. Below the threshold of $\tilde t_1 \to \tilde g \, t$, $\delta^{CP}$ remains small and $BR$ is large. The solution for an optimal measurable effect is to compromise.\\
In Fig.~\ref{fig:BR_mstop1_and_delCP_phiGluino_phiAt_only_gluino}b we show the dependence on the gluino phase $\varphi_{\tilde g}$ as a second source of CP violation. One can see the periodic behaviour of $\varphi_{\tilde g}$ and the dependence on $\varphi_{A_t}$. The $BR$ shows also a strong dependence, with higher values for higher $\varphi_{A_t}$ and $\varphi_{\tilde g} \in [\pi / 2,\pi]$.

\begin{figure}[htbp]
\begin{tabular}{cc}
\includegraphics[width=0.4\textwidth]{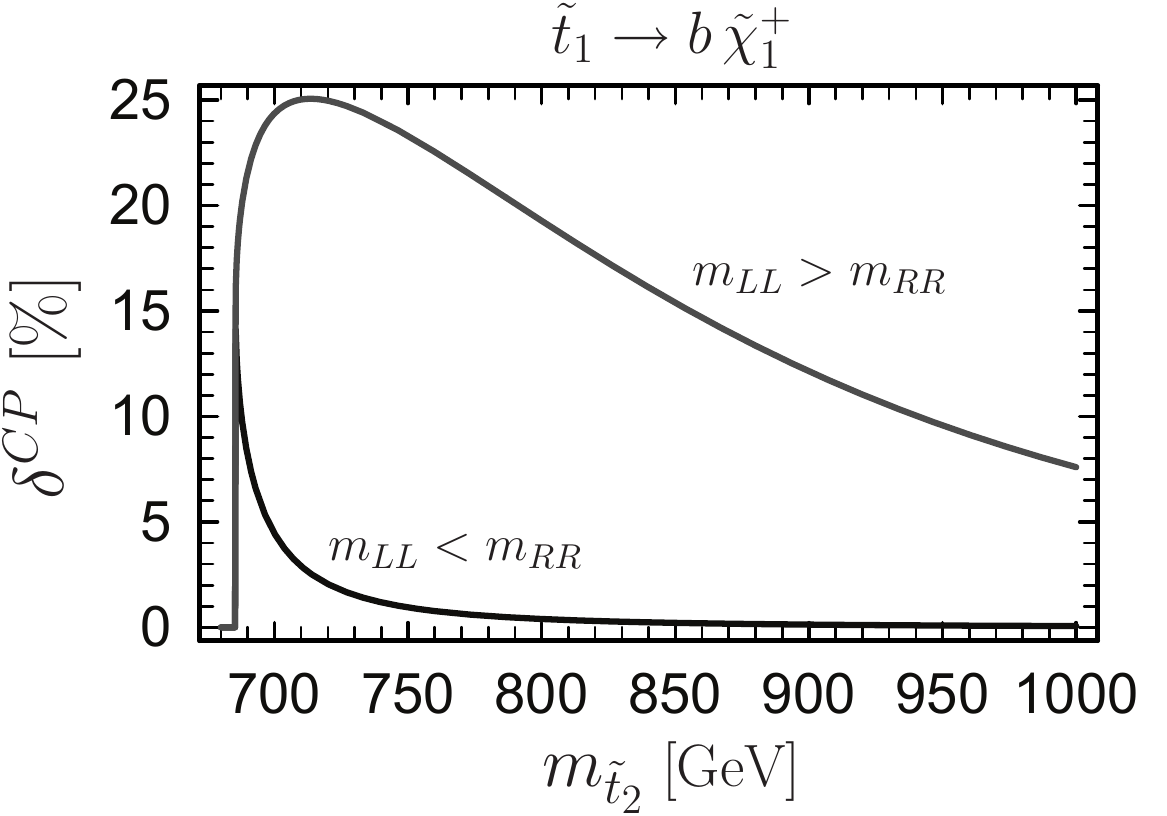} &
\includegraphics[width=0.4\textwidth]{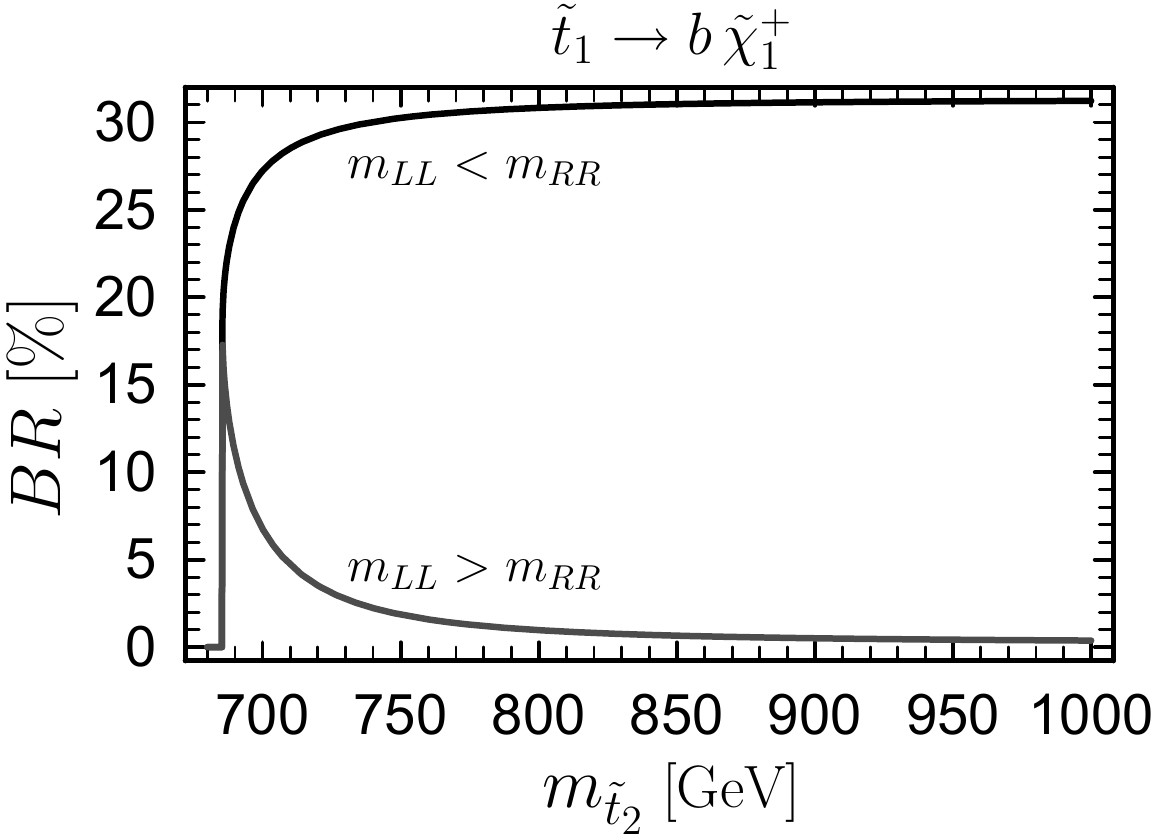}
\end{tabular}
\caption{Effect on mass splitting of $\tilde t_1$ and $\tilde t_2$ showing $\delta^{CP}$ and $BR$ as a
function of $m_{\tilde t_2}$ ($M_{\tilde Q}, M_{\tilde U}$ varied, both solutions) for $m_{\tilde t_1} = 650$~GeV.}
\label{fig:delCP&BR_mstop2_mLL_mRR}
\end{figure}

\noindent
The reason for the strong suppression of the gluino vertex graph lies in its complicated form factor, and we have not found a simple explanation yet.\\
The effect on the mass splitting of $\tilde t_1$ and $\tilde t_2$ is shown in Fig.~\ref{fig:delCP&BR_mstop2_mLL_mRR}. We fix $m_{\tilde t_1} = 650$~GeV and vary $m_{\tilde t_2}$ by changing the parameters $M_{\tilde Q}, M_{\tilde U}$. The two solutions for $M_{\tilde Q, \tilde U} (m_{\tilde t_2})$ are $m_{LL} \lessgtr m_{RR}$. The combination of $\delta_{CP}$ and $BR$ keeps low in both solutions, unless the mass splitting of $\tilde t_1$ and $\tilde t_2$ is small (due to enhancement of the $\tilde t_2$ propagator $\propto 1/(m_{\tilde t_1}^2 - m_{\tilde t_2}^2)$ in the gluino selfenergy graph).\\

\vspace*{-1cm}
\noindent
We have calculated the total cross section for $\tilde t_1$ pair production at the LHC with \textsc{Prospino}~\cite{Beenakker1998}. For $\sqrt{s}=14$~TeV, $m_{\tilde t_1}=610$~GeV and $m_{\tilde t_2}=710$~GeV we obtain $\sigma=200$~fb at NLO. Assuming ${\cal L}=300 \, \mathrm{fb}^{-1}$ (5 years), we estimate the number of CP violating events to $N = {\cal L} \times \sigma \times \delta^{CP} \times BR = 1200$ with $\delta^{CP}=0.1$ and $BR=0.2$. A possible signature is the subsequent decay $\tilde \chi^\pm_1 \to \tilde \chi^0_1 \, W^\pm \to \tilde \chi^0_1 \, l \nu_l$. Understanding the MSSM particle properties well enough, $\delta^{CP}$ can be measured at the LHC.

\section{Conclusions}

In the MSSM with complex parameters, loop corrections to the $\tilde t_i \to b \, \stilde \chi^+_k$ decay can lead to a CP violating decay rate asymmetry $\delta^{CP}$. We studied this asymmetry at full one-loop level, analyzing the dependence on the parameters and phases. A $\delta^{CP}$ of several percent is possible, mainly due to the gluino contribution in the selfenergy loop. The asymmetry can be large for low mass splitting and high mixing of the stop particles, and the chargino needs a strong wino component. But $\delta^{CP}$ must be always seen in relation to $BR$ and $\sigma_\mathrm{prod}$. The measurement should be possible already at the LHC.


\begin{theacknowledgments}
The authors acknowledge support from EU under the MRTN-CT-2006-035505 network programme. This work is supported by the "Fonds zur F\"orderung der wissenschaftlichen Forschung" of Austria, project No.~P18959-N16.
\end{theacknowledgments}






\IfFileExists{\jobname.bbl}{}
 {\typeout{}
  \typeout{******************************************}
  \typeout{** Please run "bibtex \jobname" to optain}
  \typeout{** the bibliography and then re-run LaTeX}
  \typeout{** twice to fix the references!}
  \typeout{******************************************}
  \typeout{}
 }

\end{document}